\documentclass[sigconf, ]{acmart}
\usepackage{balance}
\usepackage{booktabs} 
\usepackage{graphicx}
\usepackage{subfig}
\usepackage{color-edits}
\newcommand{\vct}[1]{\boldsymbol{#1}} 

\addauthor{pg}{red}
\addauthor{as}{blue}
\addauthor{ef}{green}
\setcopyright{rightsretained}

\acmConference[]{}{}{} 
\acmYear{2018}
\copyrightyear{2018}

\begin{document}
\title{Deep Neural Networks for Optimal Team Composition}

\author{Anna Sapienza${^*}$}
\affiliation{%
  \institution{USC Information Sciences Institute}
  \streetaddress{4676 Admiralty Way}
  \city{Marina del Rey} 
  \state{CA, USA} 
  \postcode{90292}
}
\email{annas@isi.edu}

\author{Palash Goyal${^*}$}
\thanks{*These authors contributed equally to this work.}
\affiliation{%
  \institution{USC Information Sciences Institute}
  \streetaddress{4676 Admiralty Way}
  \city{Marina del Rey} 
  \state{CA, USA} 
  \postcode{90292}
}
\email{goyal@isi.edu}

\author{Emilio Ferrara}
\affiliation{%
  \institution{USC Information Sciences Institute}
  \streetaddress{4676 Admiralty Way}
  \city{Marina del Rey} 
  \state{CA, USA} 
  \postcode{90292}
}
\email{ferrarae@isi.edu}

\begin{abstract}
Cooperation is a fundamental social mechanism, whose effects on human performance have been investigated in several environments. Online games are modern-days natural settings in which cooperation strongly affects human behavior. Every day, millions of players connect and play together in team-based games: the patterns of cooperation can either foster or hinder individual skill learning and performance. This work has three goals: (i) identifying teammates' influence on players' performance in the short and long term, (ii) designing a computational framework to recommend teammates to improve players' performance, and (iii) setting to demonstrate that such improvements can be predicted via deep learning. We leverage a large dataset from Dota 2, a popular Multiplayer Online Battle Arena game. We generate a directed co-play network, whose links' weights depict the effect of teammates on players' performance. Specifically, we propose a measure of network influence that captures skill transfer from player to player over time. We then use such framing to design a recommendation system to suggest new teammates based on a modified deep neural autoencoder and we demonstrate its state-of-the-art recommendation performance. We finally provide insights into skill transfer effects: our experimental results demonstrate that such dynamics can be predicted using deep neural networks.  
\end{abstract}

%
%



\maketitle

\section{Introduction}

Cooperation is a common mechanism  present in real world systems at various scales and in different environments, from biological organization of organisms to human society. A great amount of research has been devoted to study the effects of cooperation on human behavior and   performance~\citep{beersma2003cooperation,deutsch1960effects,johnson1989cooperation,tauer2004effects}. These works include domains spanning from cognitive learning to psychology, and cover different experimental settings (e.g., classrooms, competitive sport environments, and games), in which people were encouraged to organize and fulfill certain tasks~\citep{battistich1993interaction,cohen1994restructuring,johnson1981effects,childress2006using}. These works provide numerous insights on the positive effect that cooperation has on individual and group performance.

Many online games are examples of modern-day systems that revolve around cooperative behavior~\citep{losup2014analyzing,hudson2014measuring}. Games allow players to connect from all over the world, establish social relationships with teammates~\citep{ducheneaut2006alone}, and coordinate together to reach a common goal, while trying at the same time to compete with the aim of improving their performance as individuals. Due to their recent growth in popularity, online games have become a great instrument for experimental research. Online games provide indeed rich environments yielding plenty of contextual and temporal features related to player's behaviors as well as social connection derived from the game organization in teams.

In this work, we focus on the analysis of a particular type of online games, whose setting boosts players to collaborate to enhance their performance both as individuals and teams: Multiplayer Online Battle Arena (MOBA) games. MOBA games, such as League of Legends (LoL), Defense of the Ancient 2 (Dota 2), Heroes of the Storm, and Paragon, are examples of match-based games in which two teams of players have to cooperate to defeat the opposing team by destroying its base/headquarter. MOBA players impersonate a specific character in the battle (a.k.a., hero), which has special abilities and powers based on its role, e.g., supporting roles, action roles, etc. The cooperation of teammates in MOBA games is essential to achieve the shared goal, as shown by prior studies~\cite{yang2014identifying, drachen2014skill}. Thus, teammates might strongly influence individual players' behaviors over time.

Previous research investigated factors influencing human performance in MOBA games. On the one hand, studies focus on identifying player's choices of role, strategies as well as spatio-temporal behaviors~\citep{drachen2014skill, yang2014identifying, eggert2015classification, sapienza2017non} which drive players to success. On the other hand, performance may be affected by player's social interactions: the presence of friends~\citep{park2014social, pobiedina2013ranking}, the frequency of playing with or against certain players~\citep{losup2014analyzing}, etc. 

Despite the efforts of quantifying performance in presence of social connections, little attention has been devoted to connect the effect that teammates have in increasing or decreasing the actual player's skill level. Our study aims to fill this gap. We hypothesize that some teammates might indeed be beneficial to improve not only the strategies and actions performed but also the overall skill of a player. On the contrary, some teammates might have a negative effect on a player's skill level, e.g., they might not be collaborative and tend to obstacle the overall group actions, eventually hindering player's skill acquisition and development. 

Our aim is to study the interplay between a player's performance improvement (resp., decline), throughout matches in the presence of beneficial (resp., disadvantageous) teammates. To this aim, we build a directed co-play network, whose links exist if two players played in the same team and are weighted on the basis of the player's skill level increase/decline. Thus, this type of network only take into account the short-term influence of teammates, i.e. the influence in the matches they play together. Moreover, we devise another formulation for this weighted network to take into account possible long-term effects on player's performance.  This network incorporates the concept of ``memory'', i.e. the teammate's influence on a player persists over time, capturing temporal dynamics of skill transfer. We use these co-play networks in two ways. First, we set to quantify the structural properties of player's connections related to skill performance. Second, we build a teammate recommendation system, based on a modified deep neural network autoencoder, that is able to predict their most influential teammates.

We show through our experiments that our teammate autoencoder model is effective in capturing the structure of the co-play networks. Our evaluation demonstrates that the model significantly outperforms baselines on the tasks of (i) predicting the player's skill gain, and (ii) recommending teammates to players. Our predictions for the former result in a 9.00\% and 9.15\% improvement over reporting the average skill increase/decline, for short and long-term teammate's influence respectively. For individual teammate recommendation, the model achieves an even more significant gain of 19.50\% and 19.29\%, for short and long-term teammate's influence respectively. Furthermore, we show that using a factorization based model only marginally improves over average baseline, showcasing the necessity of non-linear models for this task.

\section{Data Collection and Preprocessing}

\paragraph{\textbf{Dota 2}} Defense of the Ancient 2 (Dota 2) is a well-known MOBA game developed and published by Valve Corporation. First released in July 2013, Dota 2 rapidly became one of the most played games on the Steam platform, accounting for millions of active players. 

We have access to a dataset of one full year of Dota 2 matches played in 2015. The dataset, acquired via \textit{OpenDota}~\citep{yasp}, consists of 3,300,146 matches for a total of 1,805,225 players. For each match, we also have access to the match metadata, including winning status, start time, and duration, as well as to the players' performance, e.g., number of kills, number of assists, number of deaths, etc., of each player.

As in most MOBA games, Dota 2 matches are divided into different categories (lobby types) depending on the game mode selected by players. As an example, players can train in the "Tutorial" lobby, or start a match with AI-controlled players in the "Co-op with AI" lobby. However, most players prefer to play with other human players, rather than with AIs. Players can decide whether the teams they form and play against shall be balanced by the player's skill levels or not, respectively in the ``Ranked matchmaking'' lobby and the ``Public matchmaking'' lobby. For Ranked matches, Dota 2 implements a matchmaking system to form balanced opposing teams. The matchmaking system tracks each player's performance throughout her/his entire career, attributing a skill level that increases after each victory and decreases after each defeat. 

For the purpose of our work, we take only into account the Ranked and Public lobby types, in order to consider exclusively matches in which 10 human players are involved. 

\begin{figure}
    \centering
    \includegraphics[width=0.8\columnwidth]{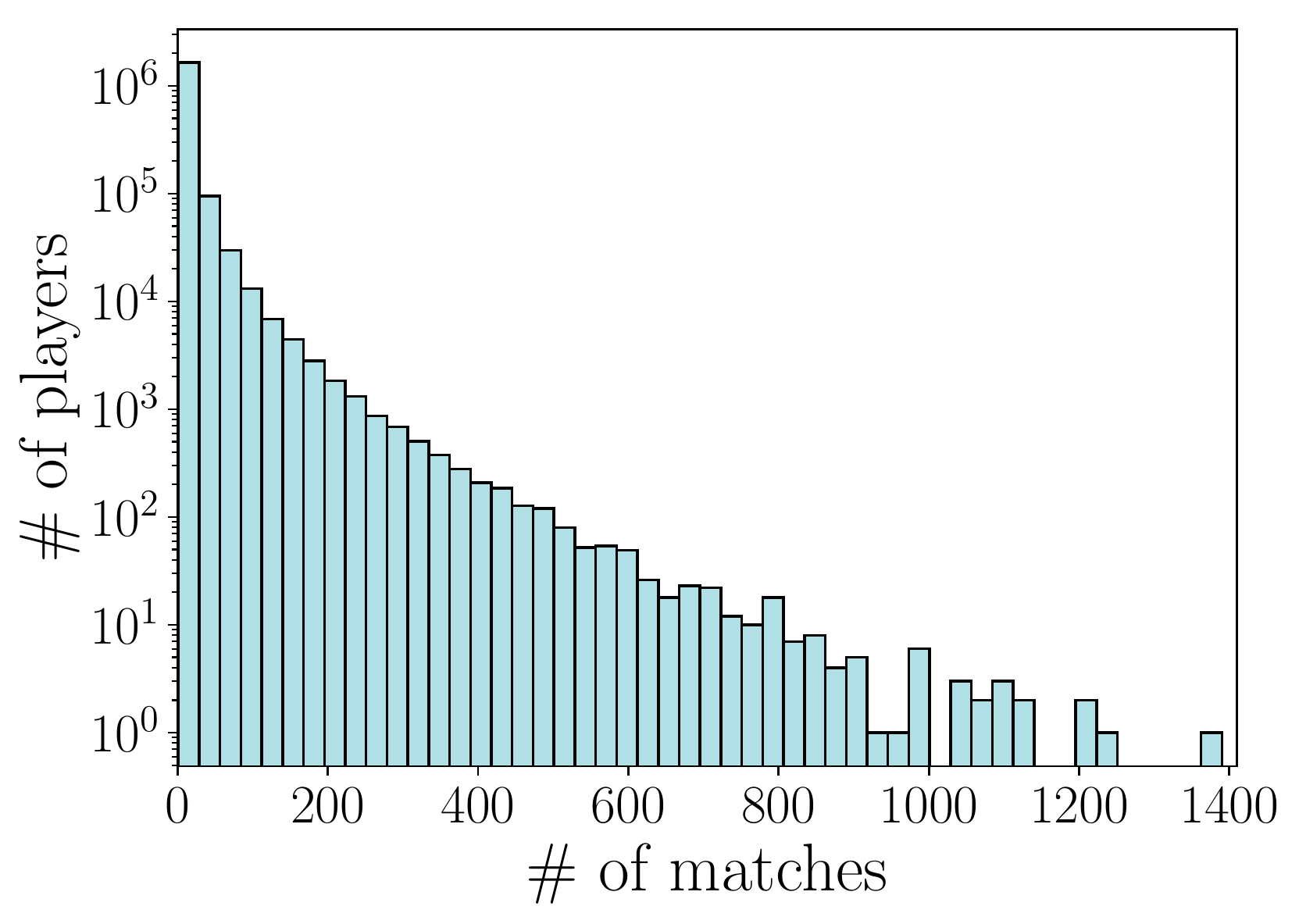}
    \vspace{-1em}
    \caption{Distribution of the number of matches per player in the Dota 2 dataset.}
    \label{fig:nmatchperpl}
    \vspace{-1.5em}
\end{figure}

\paragraph{\textbf{Preprocessing}} We preprocess the dataset in two steps. First, we select matches whose information is complete. To this aim, we first filter out matches ended early due to connection errors or players that quit at the beginning. These matches can be easily identified through the winner status (equal to a null value if a connection error occurred) and the leaver status (players that quit the game before end have leaver status equal to 0). As we can observe in Fig.~\ref{fig:nmatchperpl}, the number of matches per player has a broad distribution, having minimum and maximum values of $1$ and $1,390$ matches respectively. We note that many players are characterized by a low number of matches, either because they were new to the game at the time of data collection, or because they quit the game entirely after a limited number of matches.

In this work we are interested in assessing a teammate's influence on the skill of a player. As described in the following section, we define the skill score of a player by computing his/her TrueSkill~\cite{herbrich2007trueskill}. However, the average number of matches per player that are needed to identify the TrueSkill score in a game setting as the one of Dota 2 is 46\footnote{https://www.microsoft.com/en-us/research/project/trueskill-ranking-system/}. For the scope of this analysis, we then apply a second preprocessing step: we select all the players having at least $46$ played matches. These two filtering steps yielded a final dataset including $87,155$ experienced players.

\section{Skill Inference}

Dota 2 has an internal matchmaking ranking (MMR), which is used to track each player's level and, for those game modes requiring it, match together balanced teams. This is done with the main purpose of giving similar chance of winning to both teams. The MMR score depends both on the actual outcome of the matches (win/lose) and on the skill level of the players involved in the match (both teammates and opponents). Moreover, its standard deviation provides a level of uncertainty for each player's skill, with the uncertainty decreasing with the increasing number of player's matches. 

Player's skill is a fundamental feature that describes the overall player's performance and can thus provide a way to evaluate how each player learns and evolves over time. Despite each player having access to his/her MMR, and rankings of master players being available online, the official Dota 2 API does not disclose the MMR level of players at any time of any performed match. Provided that players' MMR levels are not available in any Dota 2 dataset (including ours), we need to reconstruct a proxy of MMR.

We overcome this issue by computing a similar skill score over the available matches: the TrueSkill~\cite{herbrich2007trueskill}. The TrueSkill ranking system has been designed by Microsoft Research for Xbox Live and it can be considered as a Bayesian extension of the well-known Elo rating system, used in chess~\cite{elo1978rating}. The TrueSkill is indeed specifically developed to compute the level of players in online games that involve more than two players in a single match, such as MOBA games. Another advantage of using such ranking system is its similarity with the Dota 2 MMR. Likewise MMR, the TrueSkill of a player is represented by two main features: the average skill of a player $\mu$ and the level of uncertainty $\sigma$ for the player's skill~\footnote{https://www.microsoft.com/en-us/research/project/trueskill-ranking-system/}. 

Here, we keep track of the TrueSkill levels of players in our dataset after every match they play. To this aim, we compute the TrueSkill by using its open access implementation in Python~\footnote{https://pypi.python.org/pypi/trueskill}. We first generate for each player a starting TrueSkill which is set to the default value in the Python library: $\mu=25$, and $\sigma=\frac{25}{3}$. Then, we update the TrueSkill of players on the basis of their matches' outcomes and teammates' levels. The resulting timelines of scores will be used in the following to compute the link weights of the co-play network.

\begin{figure}
    \centering
    \includegraphics[width=0.8\columnwidth]{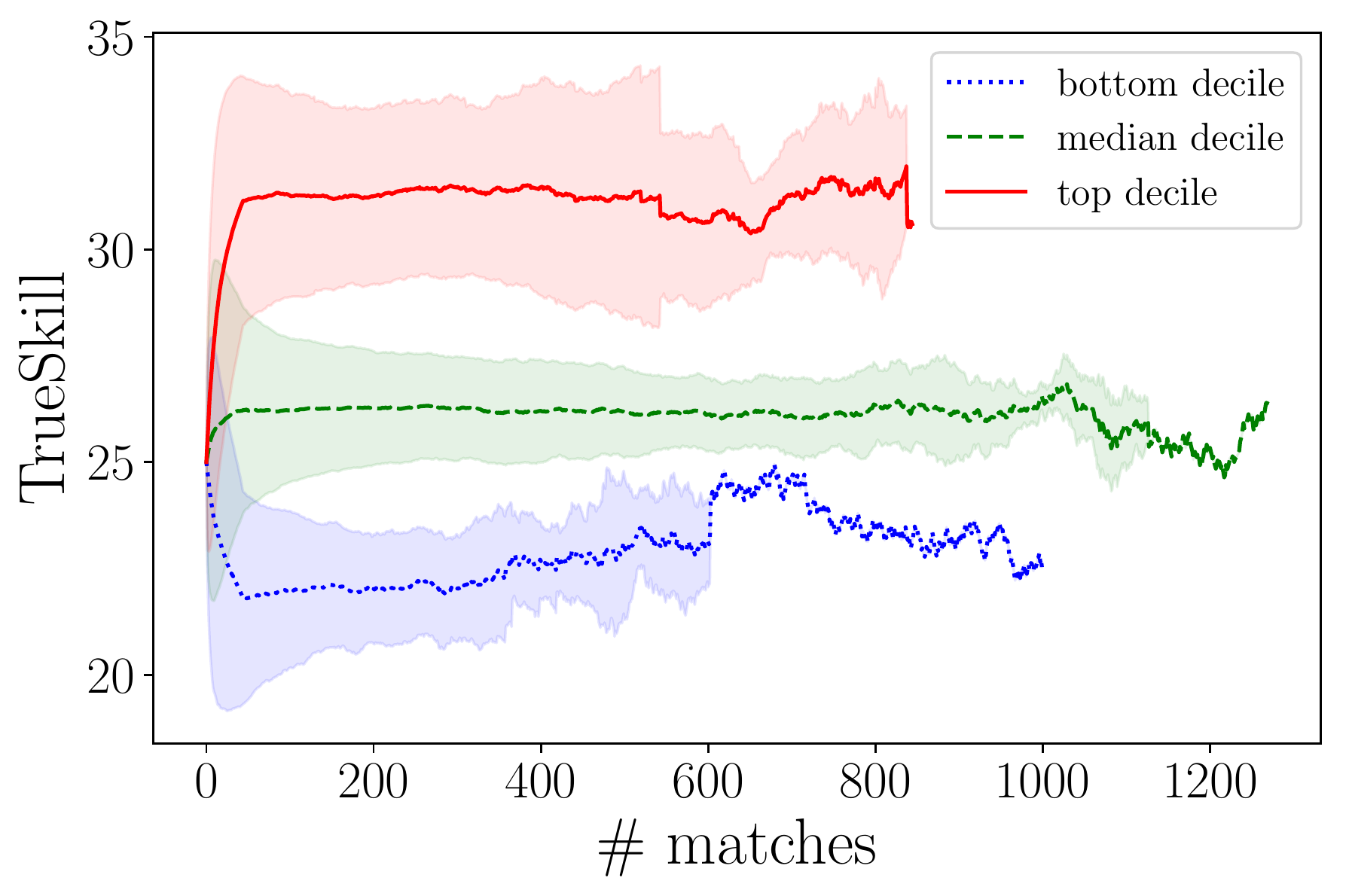}
    \vspace{-1em}
    \caption{TrueSkill timelines of players in the top, bottom, and median decile. Lines show the mean of TrueSkill values at each match index, while shades indicate the related standard deviations.}
    \label{fig:TStimelines}
    \vspace{-1.5em}
\end{figure}

For illustrative purposes, Fig.~\ref{fig:TStimelines} reports three aggregate TrueSkill timelines, for three groups of players: (i) the 10th percentile (bottom decile), (ii) the 90th percentile (top decile), and (iii) the median decile (45th-55th percentile). The red line shows the evolution of the average TrueSkill scores of the 10\% top-ranked players in Dota 2 (at the time of our data collection); the blue line tracks the evolution of the 10\% players reaching the lowest TrueSkill scores; and, the green line shows the TrueSkill progress of the ``average players''. The confidence bands (standard deviations) shrinks with increasing number of matches, showing how the TrueSkill converges with increasing observations of players' performance.\footnote{Note that the timelines have different length due to the varying number of matches played by players in each of the three deciles. In particular, in the bottom decile just one player has more than 600 matches.} The variance is larger for high TrueSkill scores. Maintaining a high rank in Dota 2 becomes increasingly more difficult: the game is designed to constantly pair players with opponents at their same skill levels, thus competition in ``Ranked matches'' becomes increasingly harsher. The resulting score timelines will be used next to compute the link weights of the co-play network. Note that, although we selected only players with at least $46$ matches, we observed timelines spanning terminal TrueSkill scores between 12 and 55. This suggests that experience alone (in terms of number of played matches) does not guarantee high TrueSkill scores, in line with prior literature~\cite{herbrich2007trueskill}.
\section{Network generation}

In the following, we explain the process to compute the co-play performance networks. In particular, we define a short-term performance network of teammates, whose links reflect TrueSkill score variations over time, and a long-term performance network, which allows to take memory mechanisms into account, based on the assumption that the influence of a teammate on a player can persist over time.

\subsection{Short-term Performance Network}

Let us consider the set of $87,155$ players in our post-processed Dota 2 dataset, and the related matches they played. For each player $p$, we define the player history as the temporally ordered set $M_p = \left[m_0,m_1,\cdots,m_N\right]$ of matches played by $p$. Each $m_i\in M_p$ is further defined to be the 4-tuple $(t_1,t_2,t_3,t_4)$ of player's teammates. Let us note that each match $m$ in the dataset can be represented as a 4-tuple because we consider just Public and Ranked matches, whose opposing teams are composed by 5 human players each. We can now define for each teammate $t$ of player $p$ in match $m_i\in M_p$ the corresponding performance weight, as:
\begin{equation}
w_{pt,m_i} = ts_{m_i}-ts_{m_{i-1}},
\end{equation}
where, $ts_{m_i}$ is the TrueSkill value of the player $p$ after match $m_i\in M_p$. Thus, weight $w_{pt,m_i}$ captures the TrueSkill gain/loss of player $p$ when playing with a given teammate $t$. This step generates as a result a time-varying directed network in which, at each time step (here the temporal dimension is defined by the sequence of matches), we have a set of directed links connecting together the players active at that time (i.e., match) to their teammates, and the relative weights based on the fluctuations of TrueSkill level of players. 

Next, we build the overall Short-term Performance Network (SPN), by aggregating the time-varying networks over the matches of each player. This network has a link between two nodes if the corresponding players were teammates at least once in the total temporal span of our dataset. Each link is then characterized by the sum of the previously computed weights. Thus, given player $p$ and any possible teammate $t$ in the network, their aggregated weight $w_{pt}$ is equal to
\begin{equation}
w_{pt} = \sum^N_{i=0} w_{pt,m_i},
\label{ws}
\end{equation}
where $w_{pt,m_i} = ts_{m_i} - ts_{m_{i-1}}$ if $t\in m_i$, and 0 otherwise. The resulting network has $87,155$ nodes and $4,906,131$ directed links with weights $w_{pt}\in\left[-0.58,1.06\right]$.

\subsection{Long-term Performance Network}

If skills transfer from player to player by means of co-play, the influence of a teammate on players should be accounted for in their future matches. We therefore would like to introduce a memory-like mechanism to model this form of influence persistence. Here we show how to generate a Long-term Performance Network (LPN) in which the persistence of influence of a certain teammate is taken into account. To this aim, we modify the weights by accumulating the discounted gain over the subsequent matches of a player as follows. Let us consider player $p$ and his/her temporally ordered sequence of matches $M_p = \left[m_0,m_1,\cdots,m_N\right]$. As previously introduced, $m_i\in M_p$ corresponds to the 4-tuple $(t_1,t_2,t_3,t_4)$ of player's teammates in that match. For each teammate $t$ of player $p$ in match $m_i\in M_p$ the long-term performance weight is defined as
\begin{equation}
w_{pt,m_i} = \exp^{i-i_{pt}}\left(ts_{m_i}-ts_{m_{i-1}}\right),
\end{equation}
where $i_{pt}$ is the index of the last match in $M_p$ in which player $p$ played with teammate $t$. Note that, if the current match $m_i$ is a match in which $p$ and $t$ play together than $i_{pt} = i$.

Analogously to the SPN construction, we then aggregate the weights over the temporal sequence of matches. Thus, the links in the aggregated network will have final weights defined by Eq.~\eqref{ws}. Conversely to the SPN, the only weights $w_{pt,m_i}$ in the LPN being equal to zero are those corresponding to all matches previous to the first one in which $p$ and $t$ co-play. The final weights of the Long-term Performance Network are $w_{pt}\in\left[-0.54,1.06\right]$.

\begin{figure}
    \centering
    \includegraphics[width=0.8\columnwidth]{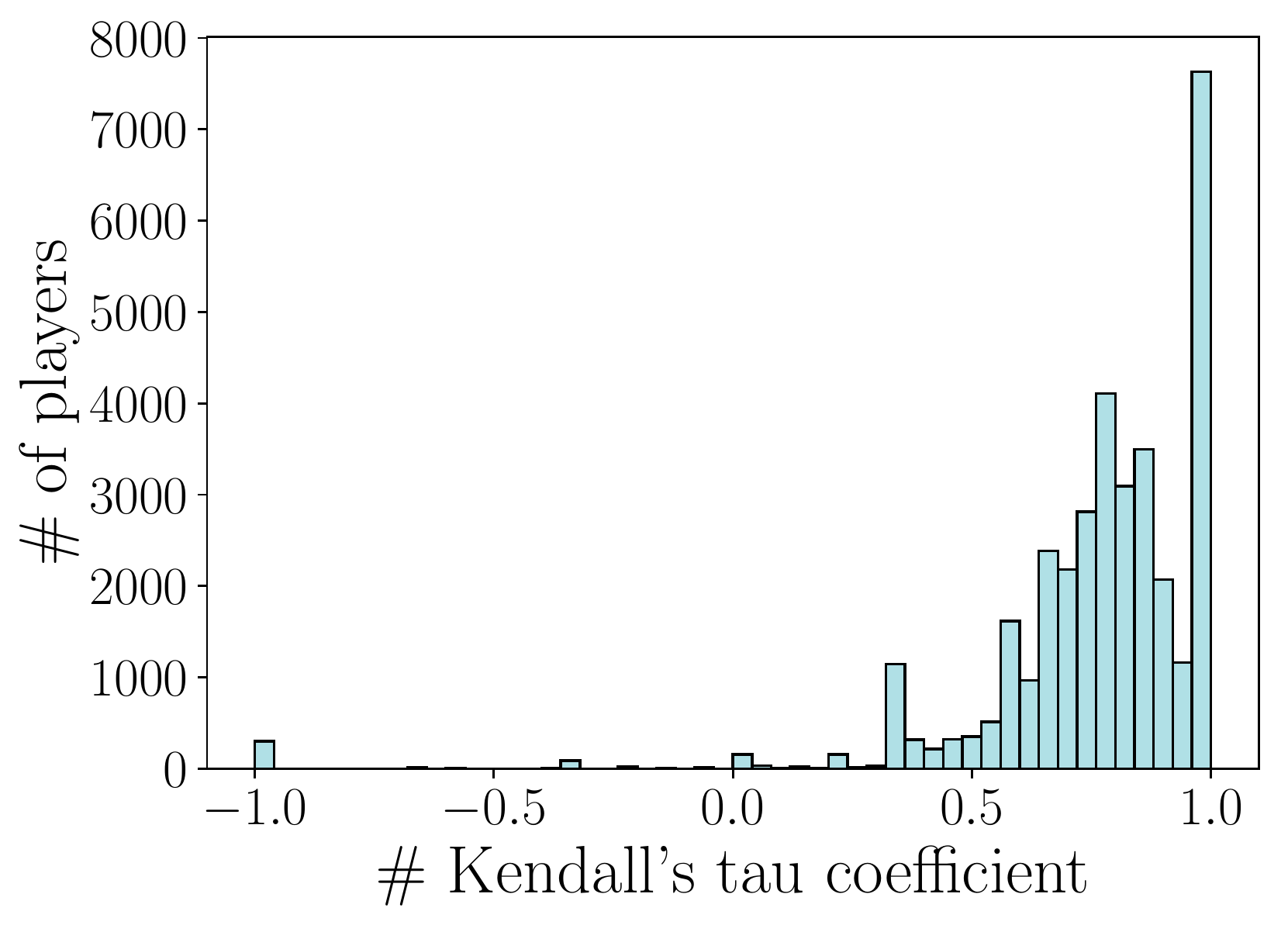}
    \vspace{-1em}
    \caption{Kendall's tau coefficient distribution computed by comparing each player's ranking in the short-term and long-term performance networks.}
    \label{fig:ktcoef}
    \vspace{-1.5em}
\end{figure}

As we can notice, the range of weights of SPN is close to the one found in LPN. However, these two weight formulations lead not only to different ranges of values but also to a different ranking of the links in the networks. When computing the Kendall's tau coefficient between the ranking of the links in the SPN and LPN, we find indeed that the two networks have a positive correlation ($\tau = 0.77$ with p-value $< 10^{-3}$) but the weights' ranking is changed. As our aim is to generate a recommending system for each player based on these weights, we further investigate the differences between the performance networks, by computing the Kendall's tau coefficient over each player's ranking. Fig.~\ref{fig:ktcoef} shows the distribution of the Kendall's tau coefficient computed by comparing each player's ranking in the SPN and LPN. In particular, we have that just a small portion of players have the same teammate's ranking in both networks, and that the $87.8\%$ of the remaining players have different rankings for their top-10 teammates. The recommending system that we are going to design will then provide a different recommendation based on the two performance networks. On the one hand, when using the SPN the system will recommend a teammate that leads to an instant skill gain. As an example, this might be the case of a teammate that is good in coordinating the team but from which not necessarily the player learns how to improve his/her performance. On the other hand, when using the LPN the system will recommend a teammate that leads to an increasing skill gain over the next matches. Thus, even if the instant skill gain with a teammate is not high, the player could learn some effective strategies and increase his/her skill gain in the successive matches.

\subsection{LCC and network properties}

\begin{figure}
    \centering
    \includegraphics[width=0.8\columnwidth]{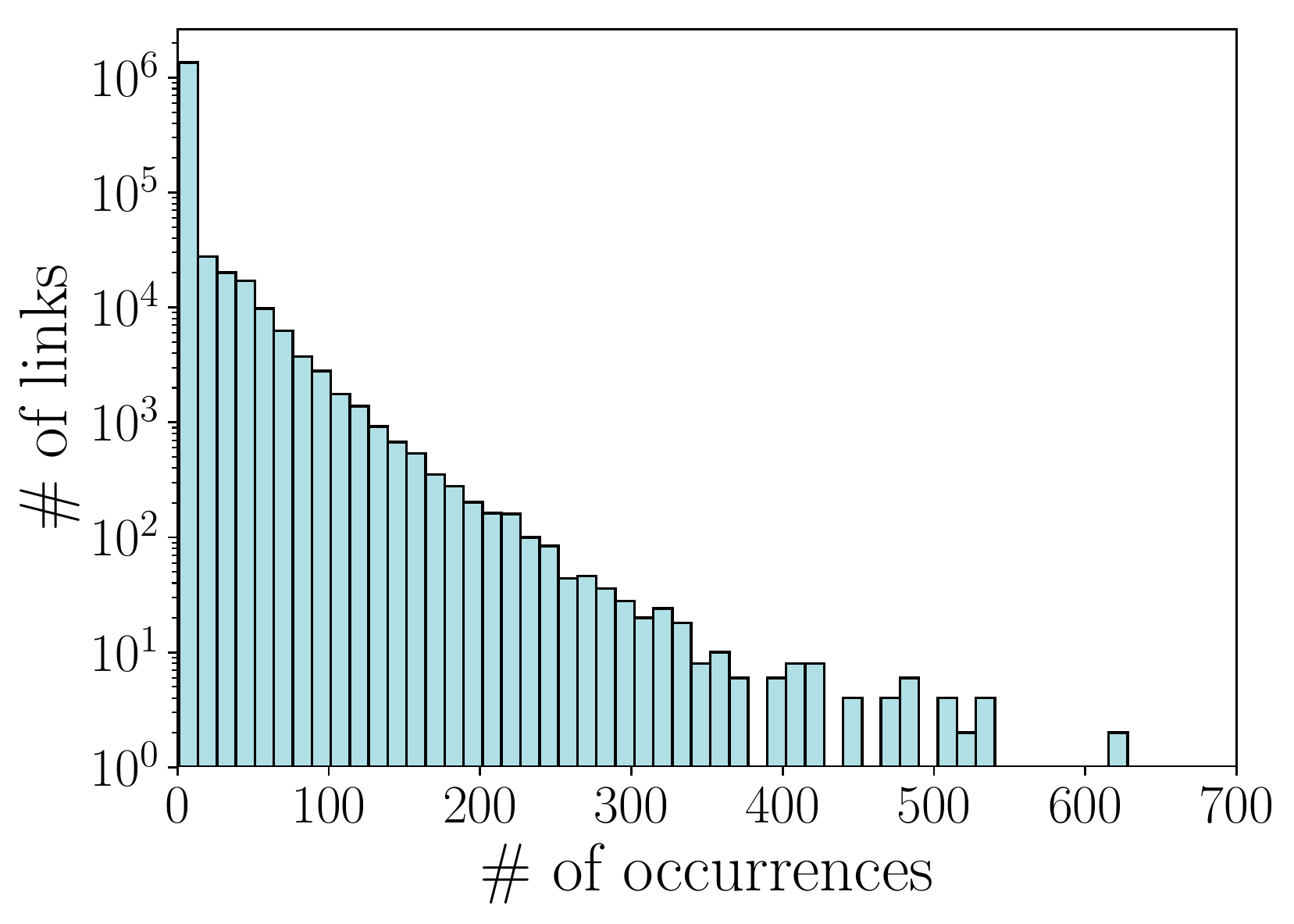}
    \vspace{-1em}
    \caption{Distribution of the number of occurrences per link, i.e. number of times a couple of teammates play together.}
    \label{fig:nocc}
    \vspace{-1.5em}
\end{figure}

Given a co-play performance network (short-term or long-term), to carry out our performance prediction we have to take into account only the links in the network having reliable weights. If two players play together just few times, the confidence we have on the corresponding weight is low. For example, if two players are teammates just one time their final weight only depends on that unique instance, and thus might lead to biased results. To face this issue, we computed the distribution of the number of occurrences a couple of teammates play together in our network (shown in Fig.~\ref{fig:nocc}) and set a threshold based on these values. In particular, we decided to retain only pairs that played more than $2$ matches together.

\begin{table}[h!]
\centering
\caption{Comparison of the overall performance networks' characteristics and its LCC. Note that the number of nodes and links are the same for both the Short-term Performance Network (SPN) and the Long-term Performance Network (LPN), while the range of weights varies from one case to the other.}
\label{nettab}
\begin{tabular}{c|c|c|c|c}
\multicolumn{1}{l|}{} & \# nodes & \# links  & SPN weights                 & LPN weights                 \\ \hline
Network               & 87,155   & 4,906,131 & $\left[-0.58, 1.06\right]$ & $\left[-0.54, 1.06\right]$  \\ \hline
LCC                   & 38,563   & 1,444,290   & $\left[-0.58, 1.06\right]$ & $\left[-0.54, 1.06\right]$
\end{tabular}
\vspace{-1.5em}
\end{table}

Finally, as many node embedding methods require a connected network as input~\cite{ahmed2017framework}, we extract the Largest Connected Component (LCC) of the performance network, which will be used for the performance prediction and evaluation. The LCC include the same number of nodes and links for both the SPN and the LPN. In particular, it includes $38,563$ nodes and $1,444,290$ links. We compare the characteristics of the initial network and its LCC in Tab.~\ref{nettab}.

\section{Performance Prediction}
In the following, we test whether the co-play performance networks have intrinsic structures allowing us to predict performance of players when matched with unknown teammates.
Such a prediction, if possible, could help us in recommending teammates to a player in a way that would maximize his/her skill improvement.

\subsection{Problem Formulation}
Consider the co-play performance network $G=(V, E)$ with weighted adjacency matrix $W$. A weighted link $(i, j, w_{ij})$ denotes that player $i$ gets a performance variation of $w_{ij}$ after playing with player $j$. We can formulate the recommendation problem as follows. Given an observed instance of a co-play performance network $G=(V, E)$ we want to predict the weight of each unobserved link $(i, j) \notin E$ and use this result to further predict the ranking of all other players $j \in V$ ($\neq i$) for each player $i \in V$.

\subsection{Network Modeling}
Does the co-play performance network contain information or patterns which can be indicative of skill gain for unseen pairs of players? If that is the case, how do we model the network structure to find such patterns? Are such patterns linear or non-linear?

To answer the above questions, we modify a deep neural network autoencoder and we test its predictive power against two classes of approaches widely applied in recommendation systems: (a) factorization based~\cite{Ahmed2013,koren2009matrix,su2009survey}, and (b) deep neural network based~\cite{Wang2016,cao2016deep,kipf2016variational}.

\subsubsection{Factorization}
In a factorization based model for directed networks, the goal is to obtain two low-dimensional matrices $U \in \mathbb{R}^{n\times d}$ and $V \in \mathbb{R}^{n\times d}$ with number of hidden dimensions $d$ such that the following function is minimized
\begin{equation*}
f(U, V) = \sum_{(i, j) \in E} (w_{ij} - <\vct{u_i}, \vct{v_j}>)^2 + \frac{\lambda}{2} (\|\vct{u_i}\|^2 + \|\vct{v_j}\|^2)
\end{equation*}
The sum is computed over the observed links to avoid of penalizing the unobserved one as overfitting to 0s would deter predictions. Here, $\lambda$ is chosen as a regularization parameter to give preference to simpler models for better generalization.

\subsubsection{Traditional Autoencoder}
\begin{figure}
    \centering
    \includegraphics[width=0.25\textwidth]{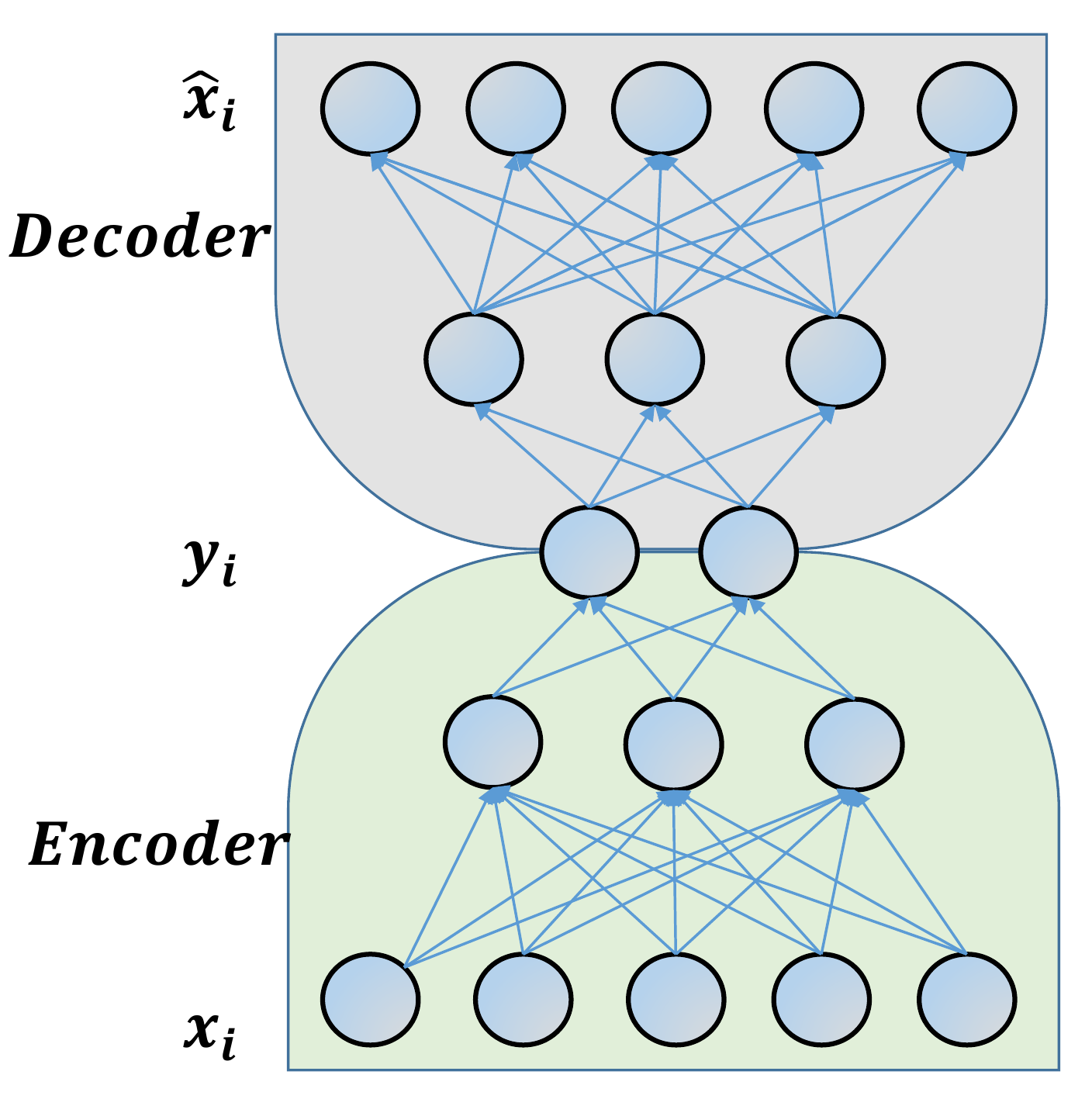}
    \vspace{-1em}
    \caption{An example of deep autoencoder model.}
    \label{fig:ae}
    \vspace{-1.5em}
\end{figure}

Autoencoders are unsupervised neural networks that aim at minimizing the loss between reconstructed and input vectors.
A traditional autoencoder is composed of two parts(cf., Figure \ref{fig:ae}): (a) an encoder, which maps the input vector into low-dimensional latent variables; and, (b) a decoder, which maps the latent variables to an output vector.
The reconstruction loss can be written as
\begin{equation}
L = \sum_{i=1}^n \|(\vct{\hat{x}_i} - \vct{x_i})\|_2^2,
\end{equation}
where $\vct{x_i}$s are the inputs and $\vct{\hat{x}_i} = f(g(\vct{x_i}))$. $f(.)$ and $g(.)$ are the decoder and encoder functions respectively. Deep autoencoders have recently been adapted to the network setting~\cite{Wang2016,cao2016deep,kipf2016variational}.
An algorithm proposed by \emph{Wang et al.}~\cite{Wang2016} jointly optimizes the autoencoder reconstruction error and Laplacian Eigenmaps~\cite{belkin2001laplacian} error to learn representation for undirected networks. However, this "Traditional Autoencoder" equally penalizes observed and unobserved links in the network, while the model adapted to the network setting cannot be applied when the network is directed. Thus, we propose to modify the Traditional Autoencoder model as follows.

\subsubsection{Teammate Autoencoder}
To model directed networks, we propose a modification of the Traditional Autoencoder model, that takes into account the adjacency matrix representing the directed network. Moreover, in this formulation we only penalize the observed links in the network, as our aim is to predict the weight and the corresponding ranking of the unobserved links. We then write our "Teammate Autoencoder" reconstruction loss as:
\begin{equation}
L = \sum_{i=1}^n \|(\vct{\hat{x}_i} - \vct{x_i}) \odot [a_{i,j}]_{j=1}^{n}\|_2^2,
\end{equation}
where $a_{ij} = 1$ if $(i, j) \in E$, and 0 otherwise.
Here, $x_i$ represents $i^{th}$ row the adjacency matrix and $n$ is the number of nodes in the network.
Minimizing this loss functions yields the neural network weights $W$ and the learned representation of the network $Y \in \mathbb{R}^{n\times d}$.

\subsection{Evaluation Framework}
\subsubsection{Experimental Setting}
\begin{figure*}[ht!]
    \centering
    \includegraphics[width=0.8\textwidth]{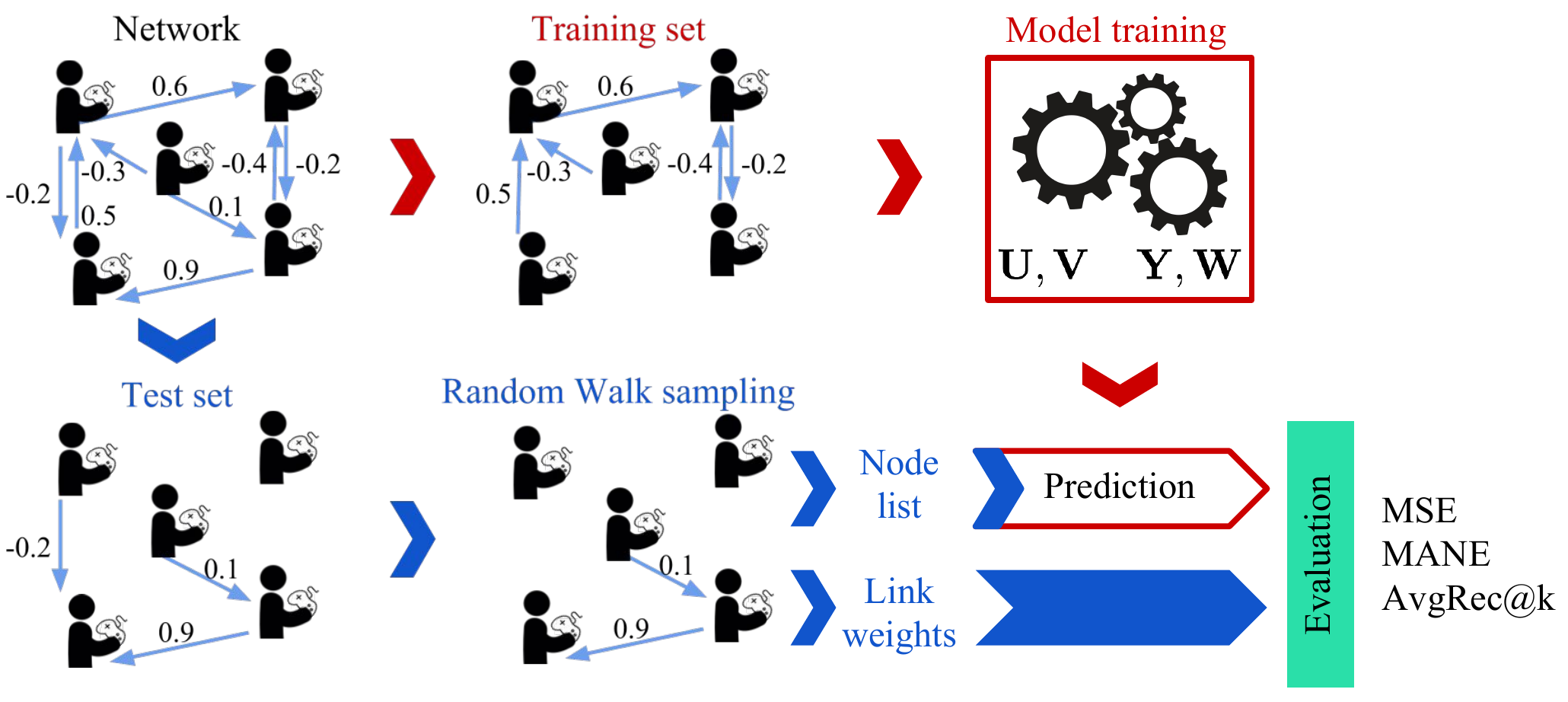}
    \vspace{-1em}
    \caption{Evaluation Framework: The co-play network is divided into training and test networks. The parameters of the models are learned using the training network. We obtain multiple test subnetworks by using a random walk sampling with random restart and input the nodes of these subnetworks to the models for prediction. The predicted weights are then evaluated against the test link weights to obtain various metrics.}
    \label{fig:schema}
    \vspace{-1.5em}
\end{figure*}
\begin{figure}
    \centering
    \includegraphics[width=0.8\columnwidth]{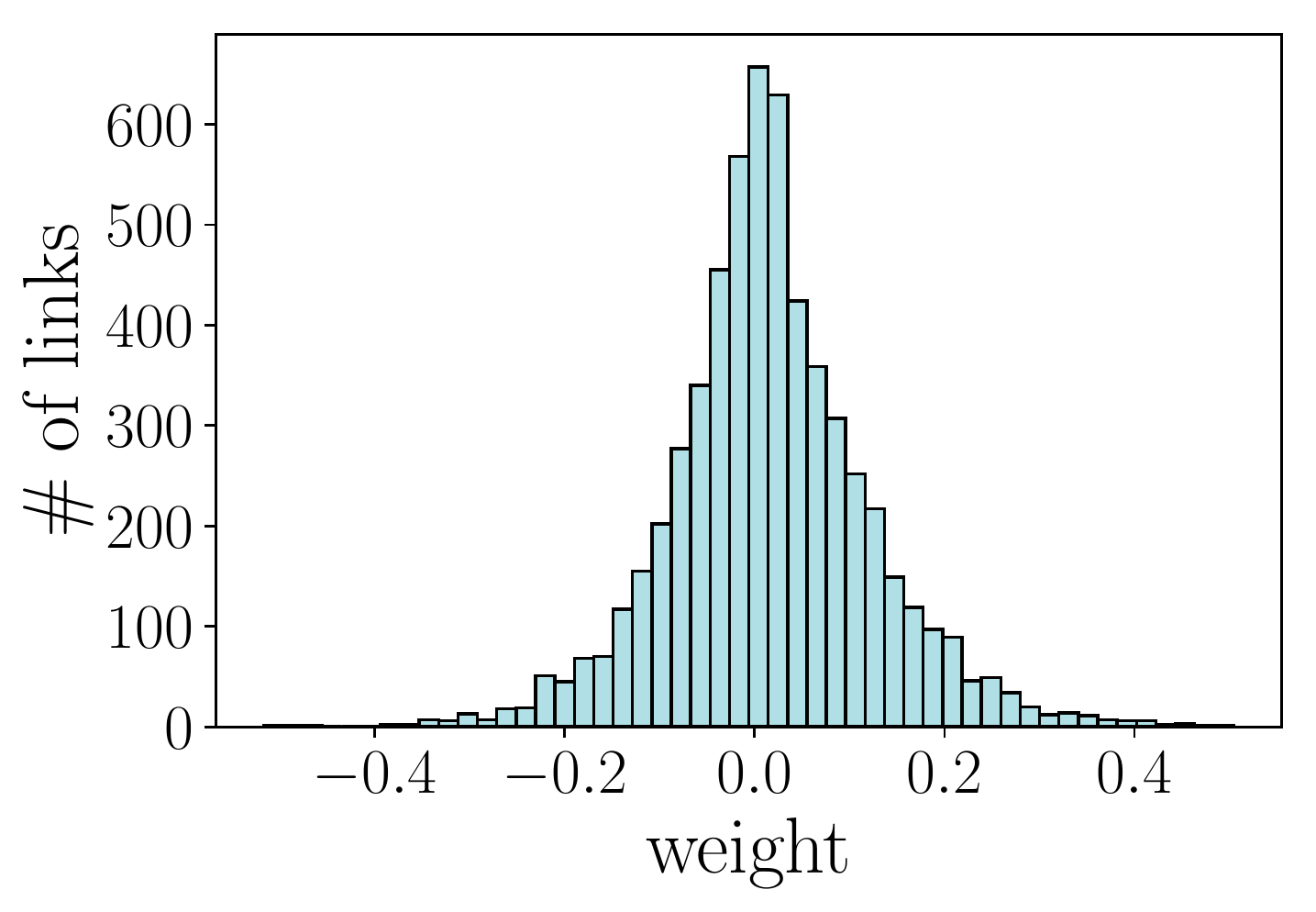}
    \vspace{-1em}
    \caption{Distribution of the weights of the network sampled by using random walk.}
    \label{fig:sampled_weightDist}
    \vspace{-2em}
\end{figure}
To evaluate the performance of the models on the task of teammates' recommendation, we use the cross-validation framework illustrated in Fig.~\ref{fig:schema}.
We randomly ``hide'' 20\% of the weighted links and use the rest of the network to learn the embedding, i.e. representation, of each player in the network.
We then use each player's embedding to predict the weights of the unobserved links. As the number of player pairs is too large, we evaluate the models on multiple samples of the co-player performance networks (similar to~\cite{Ou2016,goyal2017graph}) and report the mean and standard deviation of the used metrics.
Instead of uniformly sampling the players as performed in~\cite{Ou2016,goyal2017graph}, we use random walks~\cite{backstrom2011supervised} with random restarts to generate sampled networks with similar degree and weight distributions as the original network.
Fig.~\ref{fig:sampled_weightDist} illustrates these distributions for the sampled network of 1,024 players (nodes).

\subsubsection{Evaluation Metrics}
We use Mean Squared Error (\textit{MSE}), Mean Absolute Normalized Error (\textit{MANE}), and \textit{AvgRec@k} as evaluation metrics. \textit{MSE} evaluates the accuracy of the predicted weights, whereas \textit{MANE} and \textit{AvgRec@k} evaluate the ranking obtained by the model.

First, we compute \textit{MSE}, typically used in recommendation systems, to evaluate the error in the prediction of weights. We use the following formula for our problem:
\begin{align*}
MSE = \| \vct{w}^{test} - \vct{w}^{pred} \|^2,
\end{align*}
where $\vct{w}_{test}$ is the list of weights of links in the test subnetwork, and $\vct{w}_{pred}$ is the list of weights predicted by the model.

Second, we use \textit{AvgRec@k} to evaluate the ranking of the weights in the overall network. It is defined as:
\begin{align*}
AvgRec@k = \frac{\sum_{i=1}^k \vct{w}_{index(i)}^{test}}{k},
\end{align*}
where $index(i)$ is the index of the $i^{th}$ highest predicted link in the test network.

Finally, to test the models' recommendations for each player, we define the Mean Absolute Normalized Error (\textit{MANE}), which computes the normalized difference between predicted and actual ranking of the test links among the observed links and averages over the nodes.
Formally, it can be written as
\begin{align*}
MANE(i) &= \frac{\sum_{j=1}^{|E_i^{test}|} \left|rank_i^{pred} (j) - rank_i^{test} (j)\right|}{|E_i^{train}||E_i^{test}|},\\
MANE &= \frac{\sum_{i=1}^{|V|} MANE(i)}{|V|},
\end{align*}
where $rank_i^{pred}(j)$ represents the rank of the $j^{th}$ vertex in the list of weights predicted for the player $i$.

\subsection{Results and Analysis}
\begin{figure*}[!ht]
	\centering
	\subfloat[]{\label{fig_s1} \includegraphics[width=0.31\textwidth]{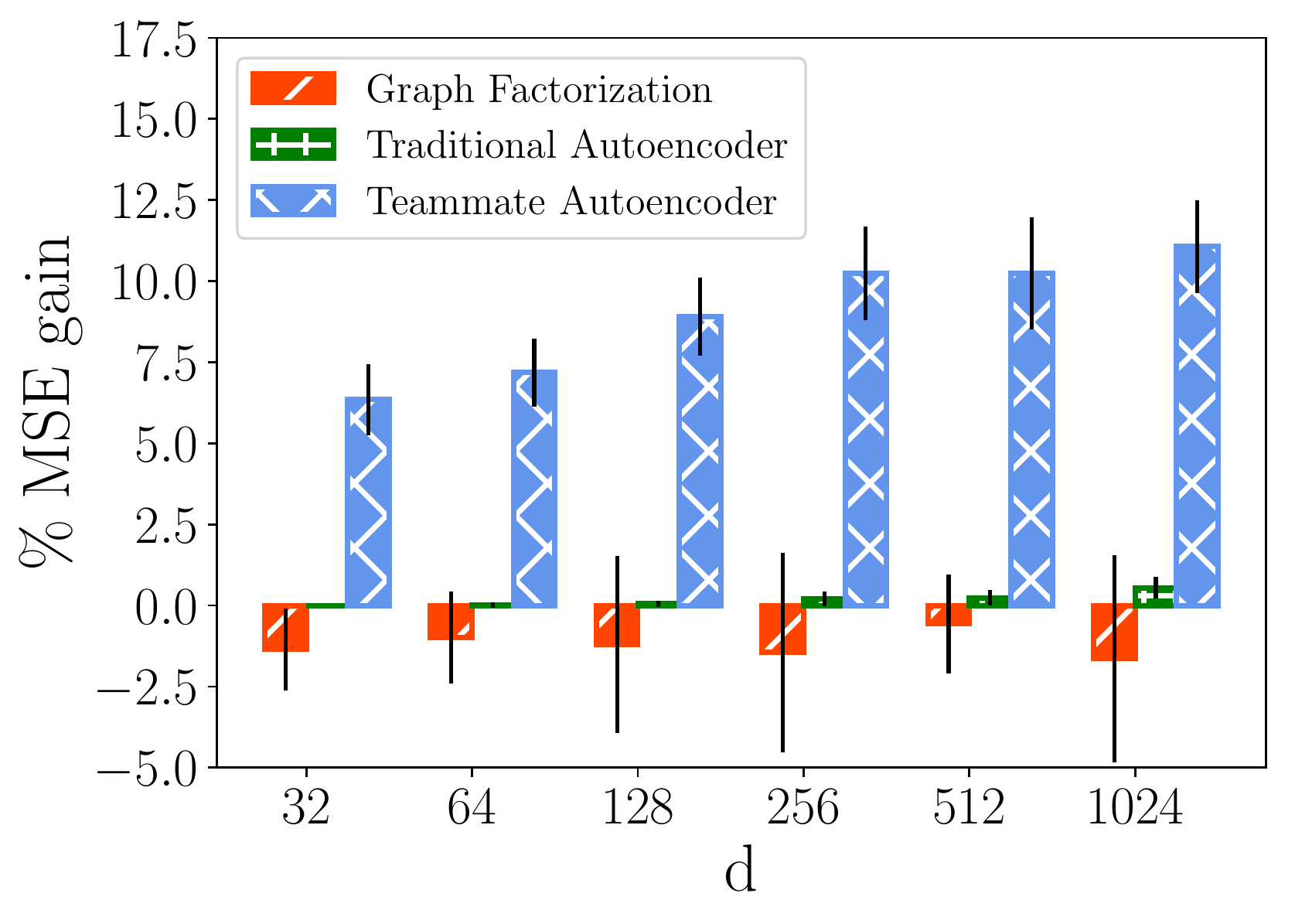}}
    \subfloat[]{\label{fig_s2} \includegraphics[width=0.31\textwidth]{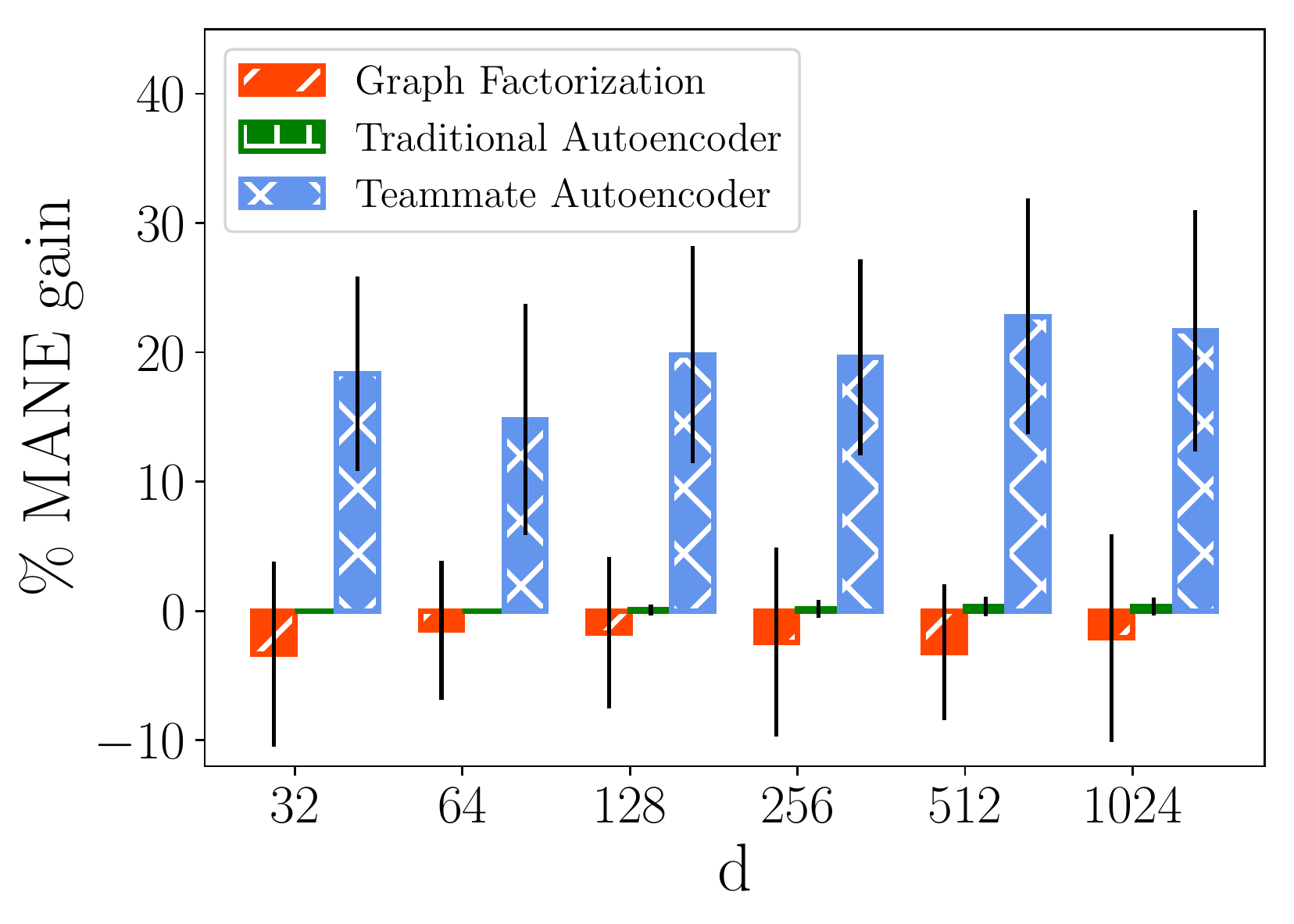}}
	\subfloat[]{\label{fig_s3} \includegraphics[width=0.31\textwidth]{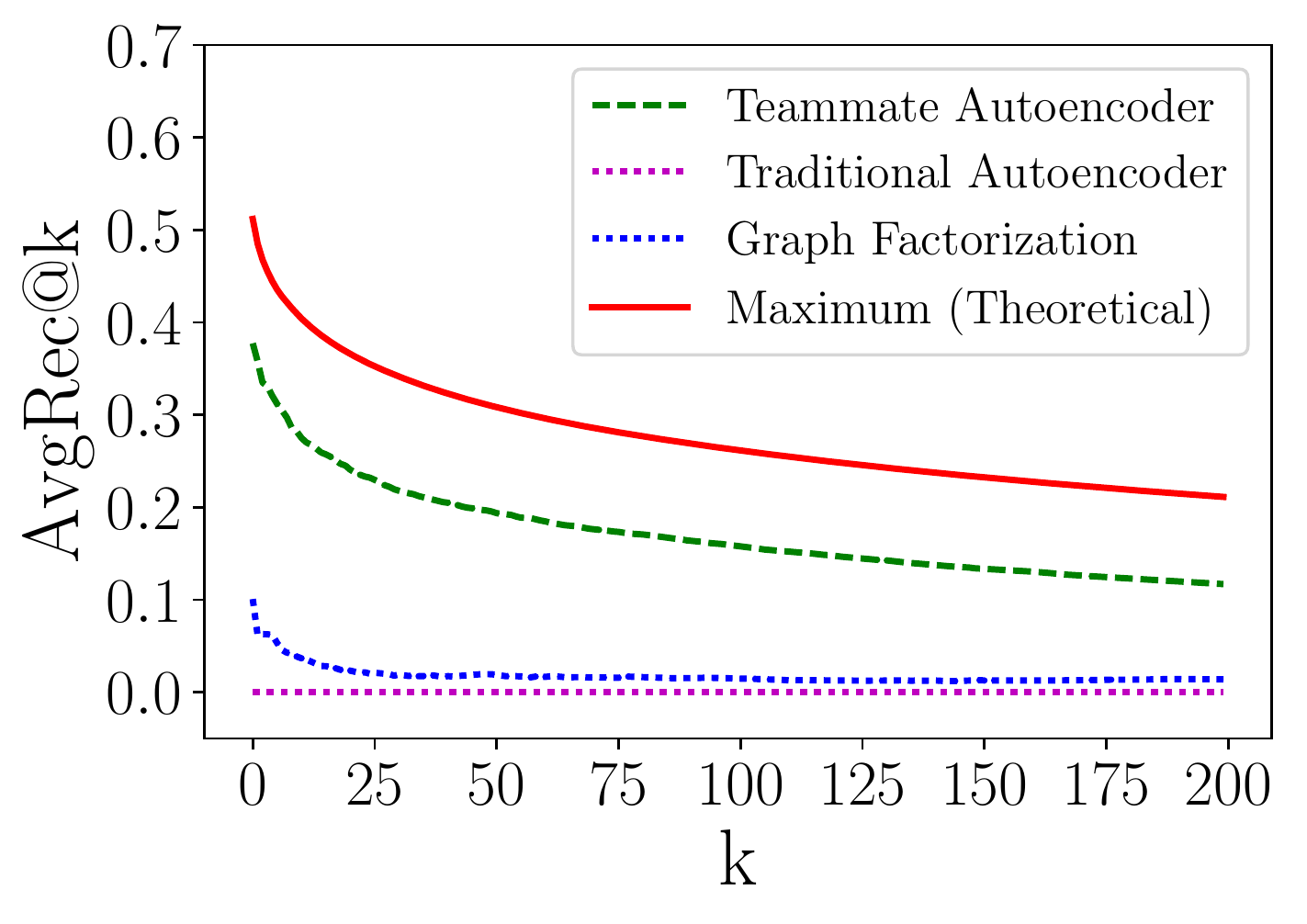}}
	\caption{Short-term Performance Network. (a) Mean Squared Error ($MSE$) gain of models over average prediction. (b) Mean Absolute Normalized Error ($MANE$) gain of models over average prediction. (c) $AvgRec@k$ of models.}
	\label{fig:mse_perf}
    \vspace{-1.5em}
\end{figure*}

\begin{figure*}[!ht]
	\centering
    \subfloat[]{\label{fig_l1} \includegraphics[width=0.31\textwidth]{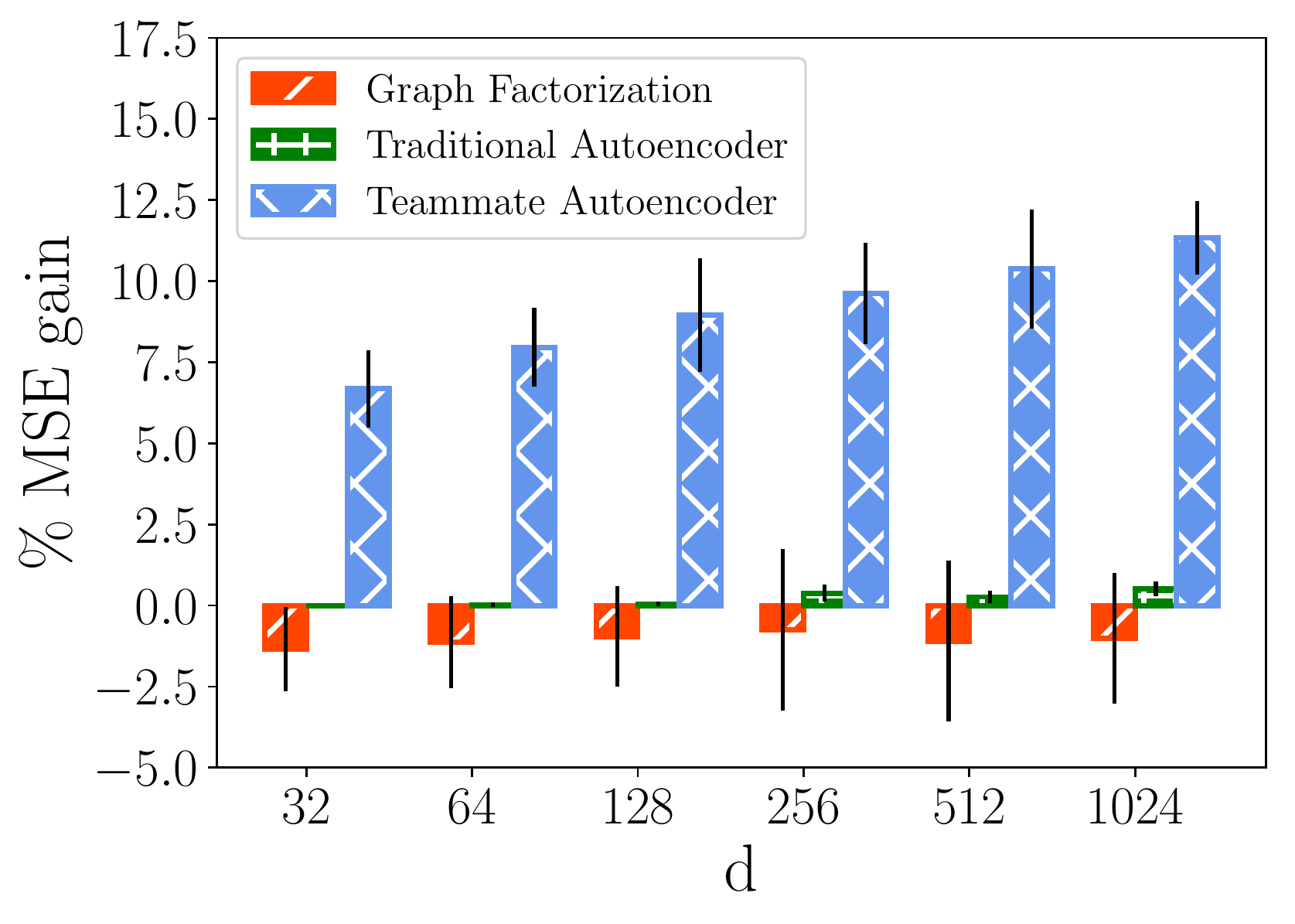}}
	\subfloat[]{\label{fig_l2} \includegraphics[width=0.31\textwidth]{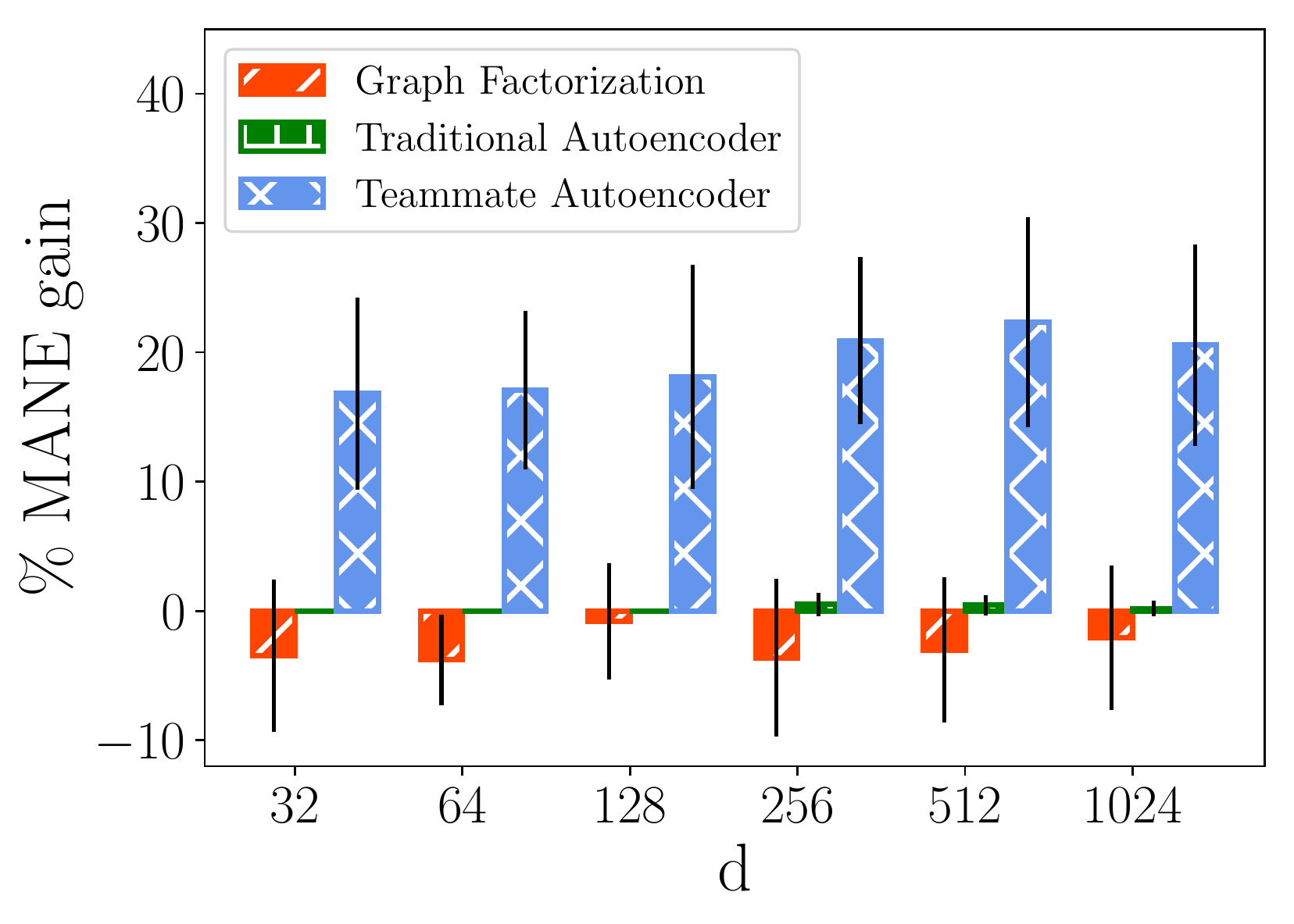}}
    \subfloat[]{\label{fig_l3} \includegraphics[width=0.31\textwidth]{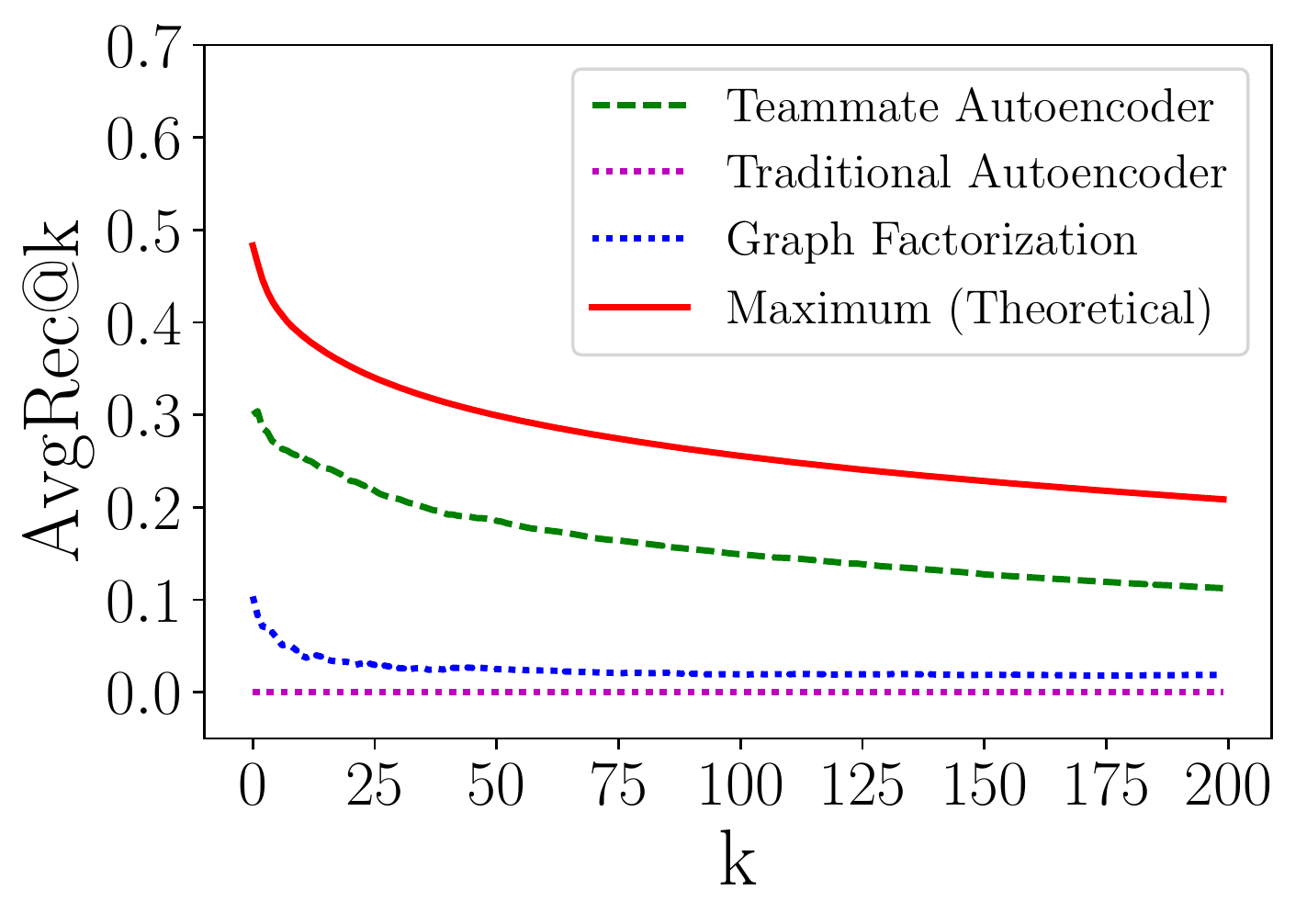}}
	\caption{Long-term Performance Network. (a) Mean Squared Error ($MSE$) gain of models over average prediction. (b) Mean Absolute Normalized Error ($MANE$) gain of models over average prediction. (c) $AvgRec@k$ of models.}
	\label{fig:mane_perf}
    \vspace{-1.5em}
\end{figure*}

\begin{table}[!ht]
\centering
\caption{Average and standard deviation of player performance prediction (\textit{MSE}) and teammate recommendation (\textit{MANE}) for $d=1,024$ in both SPN and LPN.}
\vspace{-0.8em}
\label{tab:perf_summ}
\scalebox{0.7}{
\begin{tabular}{c|c|c|c|c}
\multicolumn{1}{l|}{} & \textit{MSE}$_{SPN}$ & \textit{MANE}$_{SPN}$ & \textit{MSE}$_{LPN}$ & \textit{MANE}$_{LPN}$ \\ \hline
Baseline prediction    & 4.55/0.14   & 0.078/0.02 & 4.40/0.14 & 0.078/0.01\\ \hline
Graph Factorization   & 4.59/0.17   & 0.081/0.02 & 4.45/0.18 & 0.084/0.021\\ \hline
Traditional Autoencoder & 4.54/0.15 & 0.074/0.01 & 4.37/0.13 & 0.075/0.012 \\ \hline
Teammate Autoencoder    & \textbf{4.15/0.14} & \textbf{0.059/0.008} & \textbf{3.91/0.10} & \textbf{0.062/0.008}\\ \hline
\end{tabular}
}
\vspace{-0.8em}
\end{table}

In the following, we evaluate the results provided by the Graph Factorization, the Traditional Autoencoder and our Teammate Autoencoder. To this aim we first analyze the models' performance on both the SPN and the LPN with respect to the MSE measure, respectively in Fig.~\ref{fig_s1} and Fig.~\ref{fig_l1}. In this case, we compare the models against an ``average'' baseline, where we compute the average performance of the players' couples observed in the training set and use it as a prediction for each hidden teammate link. 

Fig.~\ref{fig_s1} and Fig.~\ref{fig_l1} show the variation of the percentage of the \textit{MSE} gain (average and standard deviation) while increasing the number of latent dimensions $d$ for in each model. We can observe that the Graph Factorization model generally performs worse than the baseline, with values in $\left[-1.64\%, -0.56\%\right]$ and average of -1.2\% for the SPN and values in $\left[-1.35\%, -0.74\%\right]$ and average of -1.05\% for the LPN. This suggests that the performance networks of Dota 2 require non-linearity to capture their underlying structure. However, a traditional non-linear model is not enough to outperform the baseline. The Traditional Autoencoder reaches indeed marginal improvements: values in $\left[0.0\%, 0.55\%\right]$ and average gain of 0.18\% for the SPN; values in $\left[0.0\%, 0.51\%\right]$ and average gain of 0.20\% for the LPN. On the contrast, our Teammate Autoencoder achieves substantial gain over the baseline across the whole spectrum and its performance in general increases for higher dimensions (they can retain more structural information). The average MSE gain for different dimensions over the baseline of the Teammate Autoencoder spans between 6.34\% and 11.06\% in the SPN and from 6.68\% to 11.34\% for the LPN, with an average gain over all dimensions of 9.00\% for the SPN and 9.15\% for the LPN. We also computed the MSE average over 10 runs and $d=1,024$, shown in Tab.~\ref{tab:perf_summ}, which decreases from the baseline prediction of 4.55 to our Teammate Autoencoder prediction of 4.15 for for the SPN, and from 4.40 to 3.91 for the LPN.

We then compare the models' performance in providing individual recommendations by analyzing the \textit{MANE} metric. Fig.~\ref{fig_s2} and Fig.~\ref{fig_l2} show the percentage of the \textit{MANE} gain for different dimensions computed against the average baseline respectively for the SPN and the LPN. Analogously to the \textit{MSE} case, the Graph Factorization performs worse than the baseline (values in $\left[-3.34\%, -1.48\%\right]$ with average gain of -2.37\% for SPN and values in $\left[-3.78\%, -0.78\%\right]$ -2.79\% for LPN) despite the increment in the number of dimensions. The Traditional Autoencoder achieves marginal gain over the baseline for dimensions higher than 128 ($\left[0.0\%,0.37\%\right]$ for SPN and $\left[0.0\%,0.5\%\right]$ for LPN), with an average gain over all dimensions of 0.16\% for SPN and 0.19\% for LPN. Our model attains instead significant percentage gain in individual recommendations over the baseline. For the SPN, it achieves an average percentage of \textit{MANE} gain spanning from 14.81\% to 22.78\%, with an overall average of 19.50\%. For the LPN, the average percentage of \textit{MANE} gain spans from 16.81\% to 22.32\%, with an overall average of 19.29\%. It is worth noting that the performance in this case does not monotonically increase with dimensions. This might imply that for individual recommendations the model overfits at higher dimensions. We report the average value of \textit{MANE} in Tab.~\ref{tab:perf_summ} for $d=1,024$. Our model obtains average values of 0.059 and 0.062, for the SPN and LPN respectively, compared to 0.078 of the average baseline for both cases. 

Finally, we compare our models against the ideal recommendation in the test subnetwork to understand how close our top recommendations are to the ground truth. To this aim, we report the \textit{AvgRec@k} metric, which computes the average weight of the top $k$ links recommended by the models. In Fig.~\ref{fig_s3} and Fig.~\ref{fig_l3}, we can observe that the Teammate Autoencoder significantly outperforms the other models, both for the SPN and LPN respectively. For the SPN, the link with the highest predicted weight by our model achieves a performance gain of 0.38 as opposed to 0.1 for Graph Factorization. This gain is close to the ideal prediction which achieves 0.52. For the LPN, instead, our model achieves a performance gain of 0.3 as opposed to 0.1 for Graph Factorization. The performance of our model remains higher for all values of $k$. This shows that the ranking of the links achieved by our model is close to the ideal ranking. Note that the Traditional Autoencoder yields poor performance on this task which signifies the importance of relative weighting of observed and unobserved links.
\section{Related Work}

There is a broad body of research focusing on online games to identify which characteristics influence different facets of human behaviors. On the one hand, this research is focused on the cognitive aspects that are triggered and affected when playing online games, including but not limited to gamer motivations to play~\citep{yee2006motivations,jansz2007appeal,choi2004people}, learning mechanisms~\citep{steinkuehler2004learning,steinkuehler2005cognition}, and player performance and acquisition of expertise~\citep{schrader2008acquisition}. On the other hand, players and their performance are classified in terms of in-game specifics, such as combat patterns~\citep{yang2014identifying,drachen2014skill}, roles~\citep{eggert2015classification,lee2015investigating}, and actions~\citep{xia2017contributes,johnson2015all,sapienza2017non}.

Aside from these different gaming features, multiplayer online games especially distinguish from other games because of their inherent cooperative design. In such games, players have not only to learn individual strategies, but also to organize and coordinate to reach better results. This intrinsic social aspect has been a focal research topic~\citep{ducheneaut2006alone,losup2014analyzing,hudson2014measuring}. In~\citep{cole2007social}, authors show that multiplayer online games provide an environment in which social interactions among players can evolve into strong friendship relationships. Moreover, the study shows how the social aspect of online gaming is a strong component for players to enjoy the game. Another study~\citep{pobiedina2013ranking,pobiedina2013successful} ranked different factors that influence player performance in MOBA games. Among these factors, the number of friends resulted to have a key role in a successful team formation. 

In the present work, we focused on social contacts at a higher level: co-play relations. Teammates, either friends or strangers, can affect other players' styles through communication, by trying to exert influence over others, etc.~\citep{kou2014playing}. Moreover, we leveraged these teammate-related effects on player performance to build a teammate recommendation system for players in Dota 2.

Recommendation systems have been widely studied in the literature on applications such as movies, music, restaurants and grocery products~\cite{lekakos2008hybrid, van2013deep, fu2014user, lawrence2001personalization}. The current work on such systems can be broadly categorized into: (i) collaborative filtering~\cite{su2009survey}, (ii) content based filtering~\cite{pazzani2007content}, and (iii) hybrid models~\cite{burke2002hybrid}. Collaborative filtering is based on the premise that users with similar interests in the past will tend to agree in the future as well. Content based models learn the similarity between users and content descriptions. Hybrid models combine the strength of both of these systems with varying hybridization strategy.

In the specific case of MOBA games, recommendation systems are mainly designed to advise players on the type of character (hero) they impersonate\footnote{Dota Picker. http://dotapicker.com/}~\citep{conley2013does,agarwala2014learning}. Few works addressed the problem of recommending teammates in MOBA games. In~\citep{van2013understanding}, authors discuss how to improve matchmaking for players based on the teammates they had in their past history. They focus on the creation and analysis of the properties of different networks in which the links are formed based on different rules, e.g., players that played together in the same match, in the same team, in adversarial teams, etc. These networks are then finally used to design a matchmaking algorithm to improve social cohesion between players. However, the author focus on different relationships to build their networks and on the strength of network links to design their algorithm, while no information about the actual player performance is taken into account. 

Here, we aim at combining both the presence of players in the same team (and the number of times they play together) and the effect that these combinations have on player performance, by looking at skill gain/loss after the game.
\section{Conclusions}
In this paper, we set to study the complex interplay between cooperation, teams and teammates' recommendation, and players' performance in online games. Our study tackled three specific problems: (i) understanding short and long-term teammates' influence on players' performance; (ii) recommending teammates with the aim of improving players skills and performance; and (iii) demonstrating a deep neural network that can predict such performance improvements.

We used Dota 2, a popular Multiplayer Online Battle Arena game hosting millions of players and matches every day, as a virtual laboratory to understand performance and influence of teammates. We used our dataset to build a co-play network of players, with weights representing a teammate's short-term influence on a player performance. We also developed a variant of this weighting algorithm that incorporates a memory mechanism, implementing the assumption that player's performance and skill improvements carry over in future games (i.e. long-term influence): influence can be intended as a longitudinal process that can improve or hinder player's performance improvement over time. 

With this framework in place, we demonstrated the feasibility of a recommendation system that suggests new teammates, which can be beneficial to a player to play with to improve their individual performance. This system, based on a modified autoencoder model, yields state-of-the-art recommendation accuracy, outperforming graph factorization techniques considered among the best in recommendation systems literature, closing the existing gap with the maximum improvement that is theoretically achievable. Our experimental results suggest that skill transfer and performance improvement can be accurately predicted with deep neural networks. 

We plan to extend this work in multiple future directions: from a theoretical standpoint, we intend to determine whether our framework can be generalized to generate recommendations and predict team and individual performance in a broader range of scenarios, beyond online games. We also plan to demonstrate, from an empirical standpoint, that the recommendations produced by our system can be implemented in real settings. We will carry out randomized-control trials in lab settings to test whether individual performance in teamwork-based tasks can be improved. One additional direction will be to extend our framework to recommend incentives alongside teammates: this to establish whether we can computationally suggest incentive-based strategies to further motivate individuals and improve their performance within teams.


\begin{acks}
The authors are grateful to DARPA for support (grant \#D16AP00115). This project does not necessarily reflect the position/policy of the Government; no official endorsement should be inferred. Approved for public release; unlimited distribution.
\end{acks}

\balance
\bibliographystyle{ACM-Reference-Format}
\bibliography{acmart.bib} 

\end{document}